\documentclass[aps,prd,twocolumn,groupedaddress,showpacs]{revtex4}

\newcommand{\ket}[1]{\ensuremath{|#1\rangle}}
\newcommand{\tE}{\tilde E}
\newcommand{\tB}{\tilde B}
\newcommand{\eps}{{\rlap{\lower2ex\hbox{$\,\,\tilde{}$}}{\epsilon_{ijk}}}}

\newcommand{\ep}{\epsilon}

\newcommand{\be}{\begin{equation}}
\newcommand{\en}{\end{equation}}
\newcommand{\cK}{\emph{K}}
\newcommand{\tv}{\tilde v}

\newcommand{\pslash}{D\kern-0.15em\raise0.17ex\llap{/}\kern0.15em\relax}
\begin{document}

\title{Further simplification of the constraints of four-dimensional gravity}
\author{Chopin Soo}\email{cpsoo@mail.ncku.edu.tw}
  \affiliation{Department of Physics, National Cheng Kung University, Tainan, Taiwan}

\begin{abstract}The super-Hamiltonian of 4-dimensional gravity as
simplified by Ashtekar through the use of gauge potential and
densitized triad variables can furthermore be succinctly expressed
as a Poisson bracket between the volume element and other
fundamental gauge-invariant elements of 3-geometry. This
observation naturally suggests a reformulation of non-perturbative
quantum gravity wherein the Wheeler-DeWitt Equation is identical
to the requirement of the vanishing of the corresponding
commutator. Moreover, this reformulation singles out spin network
states as the preeminent basis for expansion of all physical
states.
\end{abstract}

\maketitle

\section{Introduction}

Non-perturbative canonical quantization of gravity attempts to
overcome the perturbative non-renormalizability of Einstein's
theory by treating the constraints exactly. Ashtekar's seminal
simplification of the constraints through the use of gauge
connection and densitized triad variables\cite{Ash} bridged the
distinction between geometrodynamics and gauge dynamics by
identifying the densitized triad, $\tE^{ia}$ -from which the
metric is a derived composite - as the momentum conjugate to an
$SO(3,C)$ gauge potential $A_{ia}$.  The introduction of spin
network states\cite{RS-spinnetwork} have also yielded discrete
spectra for well defined area and volume operators\cite{Area-Vol}.
To the extent that exact states and rigorous results are needed,
simplifications of the classical and corresponding quantum
constraints are of great importance to the program. These include
Ashtekar's original simplification and also Thiemann's observation
that $\epsilon_{abc}\eps\tE^{ia}\tE^{jb}$ in the super-Hamiltonian
constraint is proportional to the Poisson bracket between the
connection and the volume operator\cite{Thiemann}. Thus it is not
unreasonable to expect even more progress from further
simplification of the constraints, all the more so if the
simplification is naturally associated with spin networks states.

Starting with the fundamental conjugate pair and Poisson bracket,
\be \{{\tE}^{ia}({\vec x}),A_{jb}({\vec y})\}_{P.B.} =
-i(\frac{8\pi{G}}{{c^3}})\delta^i_j \delta^a_b \delta^3({\vec x} -
{\vec y}), \label{e:abc}\en we shall show that the
super-Hamiltonian permits a further remarkable simplification: It
is in fact expressible as a Poisson bracket between the volume
element and other fundamental invariants, even when the
cosmological constant, $\lambda$, is non-vanishing. The leads
naturally to an equivalent classical constraint, and to the
quantum Wheeler-DeWitt Equation as a vanishing commutator
relation. The preeminence of spin network states come naturally
from the fact that they are eigenstates of the Hermitian volume
element operator; and can therefore be used as a basis for all
physical states.

\section{Gauge-invariant elements of 3-geometry and their Poisson Brackets}

Three physical quantities- the volume element $(\tv)$, the
Chern-Simons functional of the gauge potential ($C[A]$), and the
integral of the mean extrinsic curvature (${\cK }$)- form the
basic ingredients of 4-dimensional General Relativity as a theory
of the conjugate pair of densitized triad and gauge variables,
$(\tE^{ia}, A_{ia})$. All are gauge-invariant, but the latter two
are in addition also invariant under three-dimensional
diffeomorphisms i.e. they are elements of 3-geometry.

Their definitions are as follows: \be\tv({\vec x}) \equiv
\sqrt{\frac{1}{3!}\epsilon_{abc}\eps\tE^{ia}({\vec
x})\tE^{jb}({\vec x})\tE^{kc}({\vec x})} = |\det E_{ai}|.
\label{e:m}\en Its integral over the Cauchy surface, $M$, is the
volume, $V =\int_M \tv{({\vec x}}) d^3x$. The Chern-Simons
functional of the Ashtekar connection is \be C \equiv
\frac{1}{2}\int_M (A^a \wedge dA_a + \frac{1}{3}\ep^{abc}A_a\wedge
A_b\wedge A_c). \en Its characteristic feature is that it
satisfies $\frac{{\delta}C[A]}{\delta{A_{ia}}}={\tilde B}^{ia}$ if
$\partial M = 0$; wherein $\tB^{ia}$ is the non-Abelian $SO(3)$
magnetic field of $A_{ia}$\footnote{If $M$ is with boundary, the
imposition of appropriate boundary conditions, or the introduction
supplementary boundary terms should be considered. On the other
hand, one can treat $\partial M =0$ as a predictive element of the
present reformulation.}. The integral of the trace of the
extrinsic curvature is \be{\cK} \equiv \frac{i}{2}\int_M E^a
\wedge (D_A E)_a = \int_M ( {\tE}^{ia}k_{ia} ) d^3x,\label{e:k}\en
with the observation that the complex Ashtekar connection is
$A_{ia} \equiv -ik_{ia} + \Gamma_{ia}$, and $(D_A E)_a = dE_a
+\epsilon_{a}\,^{bc}A_b\wedge E_c$, and $\Gamma_a$ is the
torsionless connection ($dE_a + \epsilon_{ab}\,^c\Gamma_b \wedge
E_c = 0$) connection compatible with the dreibein 1-form $E_a=
E_{ai}dx^i$ on $M$.

  With the above definitions and the fundamental relation of Eq.({\ref{e:abc}}),
it follows that the following Poisson brackets hold:
\begin{eqnarray}
 \{ \tv, {\cK}
\}_{P.B.}&=&3(\frac{4\pi G}{c^3})\tv
\cr \{\tv, C \}_{P.B.} &=& (\frac{2\pi
G}{ic^3\tv}){\epsilon_{abc}\eps\tE^{ia}\tE^{jb}\tB^{kc}}\cr
\{{\cK}, C \}_{P.B.}&=&(\frac{8\pi G}{c^3})\int_M
(\tB^{ia}k_{ia})\, d^3x\cr {\tilde H} \equiv \{ \tv, iC +
\frac{\lambda}{3}{\cK}
\}_{P.B.}&=&(\frac{2\pi{G}}{c^3\tv})[\epsilon_{abc}\eps\tE^{ia}\tE^{jb}\cr
&&(\tB^{kc}+\frac{\lambda}{3}\tE^{kc})].\label{e:pb}
\end{eqnarray}

\section{Further simplification of the super-Hamiltonian constraint}

Expressed in Ashtekar variables, the super-Hamiltonian constraint
for the theory of General Relativity is {\it precisely}\cite{Ash}
\be \tilde{\tilde{H_0}} =\frac{c^3}{16\pi
G}[\epsilon_{abc}\eps\tE^{ia}\tE^{jb}(\tB^{kc} +
\frac{\lambda}{3}\tE^{kc})] \approx 0.\label{e:l}\en  It follows
that at the classical level, we may {\it equivalently} replace the
super-Hamiltonian constraint with the vanishing of a Poisson
bracket i.e. \be \{ \tv, \frac{\lambda}{3}{\cK} + iC \}_{P.B.} =
0.\en  The new super-Hamiltonian, ${\tilde H}$, is now a tensor
density of weight 1. Using $\tilde H \propto
{\tilde{\tilde{H_0}}}/{\tv}$ it can be demonstrated that the new
super-Hamiltonian constraint together with Ashtekar's
transcriptions of the Gauss' law and super-momentum constraints
remain a set of first class constraints at the classical level.

Even though we may invoke Poisson bracket-quantum commutator
correspondence $\{ ,\}_{P.B.}\mapsto (i\hbar)^{-1}[\,,\,] $, there
is no unique prescription for defining a quantum theory from its
classical correspondence. The previous observations naturally
suggest {\bf defining four-dimensional non-perturbative Quantum
General Relativity as a theory of the conjugate pair $(\tE^{ia},
A_{ia})$ with super-Hamiltonian constraint  imposed as the
vanishing commutation relation, $[\hat{\tv}({\vec x}),
\frac{\lambda}{3}{\hat{\cK}} + i{\hat C} ] = 0$}, together with
the requirement of invariance under three-dimensional
diffeomorphisms and internal gauge transformations.

\section{Reformulation of the Wheeler-DeWitt Equation, and further comments}

Physical quantum states $\ket{\Psi}$ are required to be
annihilated by the constraint : \be [ {\hat{\tv}}({\vec x}),
\frac{\lambda}{3}{\hat{\cK}} + i{\hat C} ] \ket{\Psi}
=0.\label{e:WD}\en It is very noteworthy that the Wheeler-DeWitt
Equation above is not merely symbolic, but is in fact {\it
expressed explicitly in terms of gauge-invariant 3-geometry
elements} $C$ and ${\cK}$.

The formulation is so far not confined to a particular
representation or realization of the theory. However it is most
interesting that {\it explicit realizations} and representations
of eigenstates of the volume element operator exist, and they are
precisely associated with {\it spin network states}! On a spin
network, it is known that ${\tv}^2$ acts in a well-defined
manner\cite{Area-Vol}. Its eigenstates are linear combinations of
spin network states of the same vertex valency(number of links) at
${\vec x}$, such that \be [\hat{\tv}({\vec x})]^2\ket{\Psi_{v}} =
v^2\ket{\Psi_{v}},\en has spectrum given by $v = 0$ if valency of
the vertex at ${\vec x}$ is less than 4, and $v$ can be computed
and may be non-trivial if the valency of the vertex at ${\vec x}$
is equal to or greater than four\cite{Area-Vol}. In the
connection-representation of spin network states, the wave
function, $\langle A\ket{\Psi}=\langle A\ket{{\bf\Gamma},
\{vertices\}, \{j\}}$, is such that for the spin network,
${\bf\Gamma}$, each link between two vertices at ${\vec x}$ and
${\vec y}$ is associated with a non-integrable phase factor ${P[
\exp(i{\int^{\vec y}_{\vec x}}_p \,A \,)]}$ with $A$ in the
spin-$j$ representation of the Lie algebra. With appropriate
assignments of combinations of Wigner symbols at the vertices,
spin network states are gauge-invariant.

Simultaneous eigenstates, $\ket{\Psi_{v,\gamma}}$, with
eigenvalues $(v,\gamma)$  of the volume element operator and the
(dimensionless) operator $\hat{\Upsilon} \equiv
\frac{\lambda}{3}\hat{\cK} + i{\hat C}$ are solutions. But since
$\hat{\Upsilon}$ does not commute off-shell with the volume
element, well defined non-trivial simultaneous eigenstates may not
exist.

The operator $\hat{\cK}$ has in fact been studied by Borrisov, De
Pietri and Rovelli\cite{BDR}, and its action on loop or
non-integrable phase factor elements of spin network states can be
made well-defined. This indicates the action of the operator
$\hat{\Upsilon} \equiv \frac{\lambda}{3}\hat{\cK} + iC$ can also
be defined on spin network states. The physical meaning of
$\hat{\cK}$ is not readily apparent in the spin network
formulation, but its connection to ``intrinsic time" in quantum
gravity may nevertheless be deduced from a different perspective.
Without resorting to particular representations, it is readily
verified that, apart from a multiplicative constant, ${\cK}$  is
in fact conjugate to the intrinsic time variable i.e. $\{\ln\tv,
{\cK} \}_{P.B.}= \frac{12\pi G}{c^3}$. In the quantum context
$\hat{\cK}$ is thus proportional to the {\it generator of
translations} of $\ln\tv = \ln|\det{E_{ai}}|$ which is furthermore
a monotonic function of the superspace ``intrinsic time variable"
($\propto \sqrt{|\det E_{ai}|}$) discovered by DeWitt in his
seminal study of canonical quantum gravity\cite{DeWitt}. The
operator ${\cK}$ is thus a Schwinger-Tomonaga ``time-evolution
operator" (complete with $i$) for an intrinsic time variable.

A few observations on the physical requirements of the inner
product and the properties of the operators with respect to
Hermitian conjugation are in order. At the classical level ${\tv}$
and $\cK$ are real(according to Eqs. ({\ref{e:m}}) and
({\ref{e:k}})), and it is reasonable to require that the inner
product must lead to Hermitian $\hat{{\tv}}$ and ${\hat{\cK}}$.
The Ashtekar connection on $M$ is furthermore understood to be the
pullback of the self-dual projection of the spin connection to the
Cauchy surface. Thus it is also reasonable to conclude that
$A^\dagger$ corresponds to the anti-self-dual projection, or the
orientation-reversed transform, of $A$. In fact these observations
suggest that $C^\dagger$ is the Chern-Simons functional for the
Ashtekar connection of the manifold ${\bar M}$ with the reversed
orientation with respect to $M$. Thus the operator
${\hat{\Upsilon}}$ is not in general P and CP-invariant. In this
respect, spin network states are holomorphic in $A$ and are
therefore also not automatically P-invariant. Further discussions
on possible P, CP and CPT violations in Ashtekar theory coupled to
matter have been discussed elsewhere\cite{chang}.

   The observations presented here highlight many remarkable features in four-dimensional
   Quantum General Relativity and behoove further consideration and continued
research.

\section*{Acknowledgments}
This work has been supported by funds from the National Science
Council of Taiwan under Grant Nos. NSC93-2112-M-006-011 and
NSC94-2112-M-006-006.


\begin{thebibliography}{}
\bibitem{Ash}
A. Ashtekar, {\it Phys. Rev. Lett.} {\bf 57}, 2244 (1986); A.
Ashtekar, {\it Phys. Rev. D}{\bf 36}, 1587 (1987); A. Ashtekar,
{\it Lectures on non-perturbative canonical gravity} (World
Scientific, 1991).

\bibitem{RS-spinnetwork}
C. Rovelli and L. Smolin, {\it Phys. Rev. D}{\bf 52}, 5743 (1995).


\bibitem{Area-Vol}
C. Rovelli and L. Smolin, {\it Nucl. Phys. B}{\bf 442}, 593
(1995), erratum-{\it ibid.} B{\bf 456}, 753 (1995); R. De Pietri
and C. Rovelli, {\it Phys. Rev. D}{\bf 54} 2664 (1996); A.
Ashtekar and J. Lewandowski, {\it Class. Quantum Grav.} {\bf 14},
A55 (1997); J. Lewandowski, {\it Class. Quant. Grav.} {\bf 14}, 71
(1997); A. Ashtekar and J. Lewandowski, {\it  Adv. Theor. Math.
Phys.} {\bf 1}, 388 (1998); J. Brunnemann and T. Thiemann, {\it
Simplification of the spectral analysis of the volume operator in
loop quantum gravity}, gr-qc/0405060; K. A. Meissner, {\it
Eigenvalues of the volume operator in loop quantum gravity},
gr-qc/0509049.

\bibitem{Thiemann}T. Thiemann, {\it Phys. Lett. B}{\bf 380}, 257 (1996).


\bibitem{BDR}R. Borissov, R. De Pietri
and C. Rovelli, {\it Class. Quantum Grav.} {\bf 14}, 2793 (1997).

\bibitem{DeWitt}Bryce S. DeWitt, {\it Phys. Rev.} {\bf 160}, 1113 (1967).

\bibitem{chang}
L. N. Chang and C. Soo, {\it Phys. Rev. D}{\bf 53}, 5682 (1996);
C. Soo and L. N. Chang, {\it The Weyl theory of fundamental
interactions: is CPT violated ?}, hep-th/9702171; L. N. Chang and
C. Soo, {\it Class. Quantum Grav.} {\bf 20}, 1379 (2003).

\end{thebibliography}
\end{document}